\documentclass[twocolumn,prb]{revtex4}
\usepackage{amsfonts}
\usepackage[T1]{fontenc}
\usepackage{amsmath,amsbsy,amssymb,graphicx}
\usepackage{times}

\begin{document}

\title{ Topological Semimetals carrying Arbitrary Hopf Numbers: \\
Hopf-Link, Solomon's-Knot, Trefoil-Knot and Other Semimetals}
\author{Motohiko Ezawa}
\affiliation{Department of Applied Physics, University of Tokyo, Hongo 7-3-1, 113-8656,
Japan}

\begin{abstract}
We propose a new type of Hopf semimetals indexed by a pair of numbers $(p,q)$%
, where the Hopf number is given by $pq$. The Fermi surface is given by the
preimage of the Hopf map, which consists of loops nontrivially linked for a
nonzero Hopf number. The Fermi surface forms a torus link, whose examples
are the Hopf link indexed by $(1,1)$, the Solomon's knot $(2,1)$, the double
Hopf-link $(2,2)$ and the double trefoil-knot $(3,2)$. We may choose $p$ or $%
q$ to be a half integer, where the Fermi surface is a torus knot such as the
trefoil knot $(3/2,1)$. It is even possible to make the Hopf number an
arbitrary rational number, where a semimetal whose Fermi surface forms open
strings is generated.
\end{abstract}

\maketitle

\textit{Introduction:} Weyl semimetals are described by the two-band model
equipped with a point node in the three-dimensional (3D) space\cite%
{Hosur,Jia}. It is characterized by the monopole charge in the momentum space%
\cite{Murakami}. Line-nodal semimetals or loop-nodal semimetals are also
possible in the 3D space, where the zero-energy Fermi surface is given by a
line or a closed loop\cite%
{Mandal,Burkov,Philip,Xie,Yu,Kim,Yamakage,Hyper,CFang,Hirayama,Sy}.
Recently, a nodal-chain semimetal is proposed, where loop nodes touch with
each other\cite{Chain}. A natural question is whether nontrivial
Fermi-surfaces made of loop nodes such as links and knots are possible.

The two-band Hamiltonian $H=\mathbf{S}\left( \mathbf{k}\right) \cdot \mathbf{%
\sigma }$ is a prototype of Hopf insulators\cite%
{Moore,Deng,Duan,DengC,Ken,Xu}, where $\mathbf{\sigma }=(\sigma _{x},\sigma
_{y},\sigma _{z})$ are the Pauli matrices and $\mathbf{S}\left( \mathbf{k}%
\right) $ is the normalized pseudospin $|\mathbf{S}\left( \mathbf{k}\right)
|=1$ spanning the sphere surface $\mathbb{S}^{2}$. On the other hand, the 3D
Brillouin zone is identical to the torus $\mathbb{T}^{3}$. A homotopy from $%
\mathbb{T}^{3}$ to $\mathbb{S}^{2}$ is characterized by the Hopf number. It
has been argued that Hopf insulators with arbitrary Hopf numbers are possible%
\cite{Duan}. Nontrivial Hopf textures are also discussed in cold atoms\cite%
{Kawaguchi, Hall}, light fields\cite{Kedia} and liquid crystal\cite{Ak}.

\begin{figure}[t]
\centerline{\includegraphics[width=0.45\textwidth]{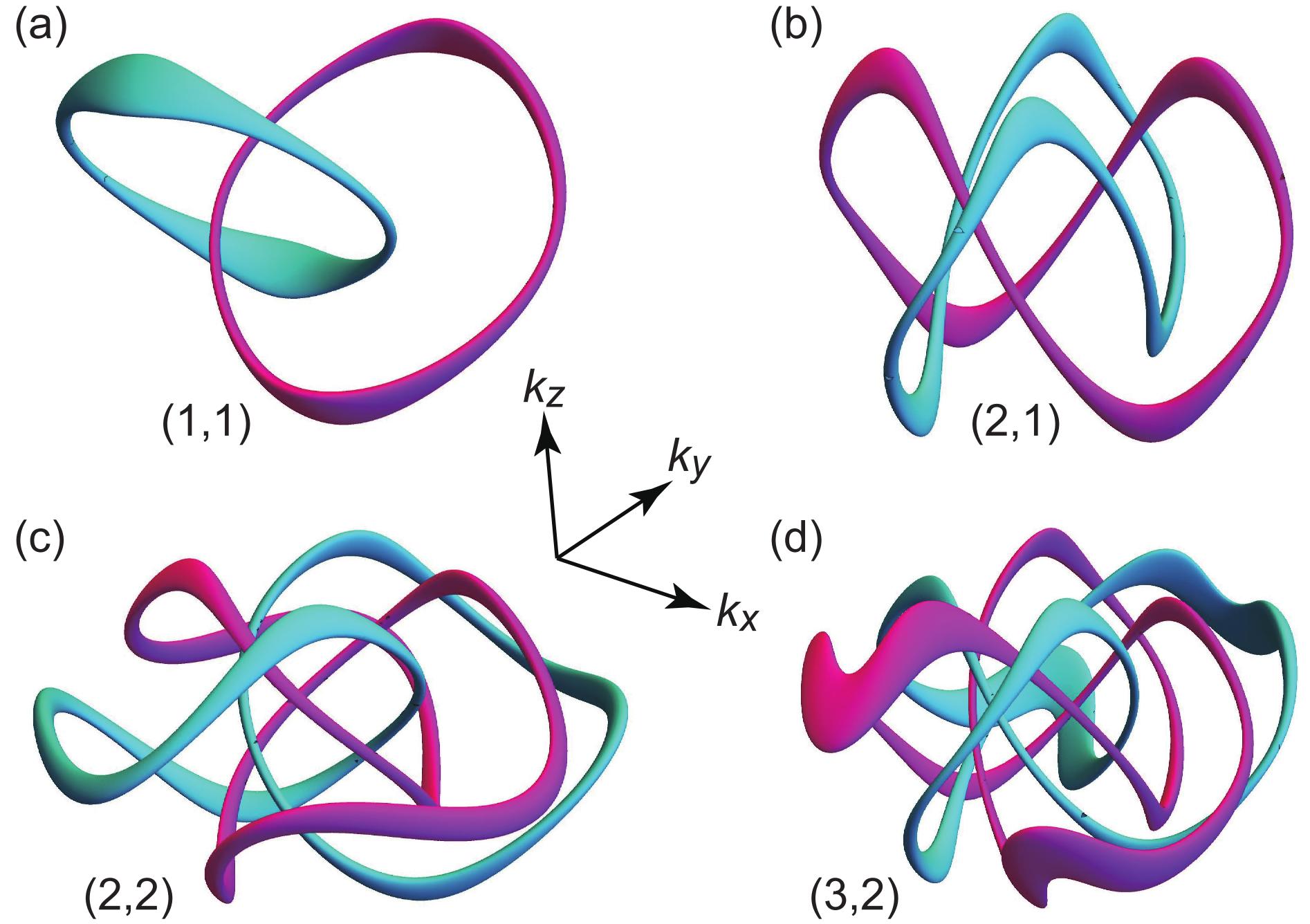}}
\caption{ Bird's eye's view of the almost zero-energy surface of the
Hamiltonian $H_{xz}$. (a) The Hopf-link semimetal with $(p,q)=(1,1)$, (b)
the Solomon's-knot semimetal with $(2,1)$, (c) the double Hopf-link
semimetal with $(2,2)$, which consists of two Hopf links. (d) the double
trefoil-knot semimetal with $(3,2)$. We have chosen $m=2$ in Eq.(\protect\ref%
{eta}) to draw figures. The preimage of $S_{y}=1$ is colored in magenta,
while that of $S_{y}=-1$ is colored in cyan.}
\label{FigLink}
\end{figure}

In this Letter we investigate topological semimetals, where Fermi surfaces
consist of nontrivial loops with arbitrary Hopf numbers. We explore the
Hamiltonian $H=S_{x}\sigma _{x}+S_{z}\sigma _{z}$, where the zero-energy
condition reads $S_{y}=\pm 1$. The Fermi surface is the preimage of the
points $\mathbf{S}_{\pm }=(0,\pm 1,0)$ in the mapping $\mathbb{T}%
^{3}\rightarrow \mathbb{S}^{2}$, and it consists of two loops. They are
linked for a nonzero Hopf number. We construct Fermi surfaces comprised of
the Hopf link, the Solomon's knot and others: See Fig.\ref{FigLink}. They
are "torus links" lying on the surface of a torus. Furthermore, we construct
Fermi surfaces comprised of torus knots by choosing half integer Hopf
numbers, among which there arises in particular a trefoil-node Fermi
surface. We can even choose any one rational number as a Hopf number, where
the Fermi surface forms open strings though it describes no longer a
topological semimetal.

\textit{Torus-link semimetals:} The Hamiltonian of topological Hopf
insulators is given by\cite{Moore,Deng,Duan,DengC,Ken,Xu}%
\begin{equation}
H\left( \mathbf{k}\right) =\mathbf{S}\left( \mathbf{k}\right) \cdot \mathbf{%
\sigma },  \label{HamilTH}
\end{equation}%
where $\mathbf{S}\left( \mathbf{k}\right) $ is the normalized pseudospin
field, $|\mathbf{S}\left( \mathbf{k}\right) |=1$. It is written as $\mathbf{S%
}=\mathbf{z}^{\dagger }\mathbf{\sigma z}$ in terms of the CP$^{1}$ field $%
\mathbf{z}$. We define it by%
\begin{equation}
\mathbf{z}\equiv \left( 
\begin{array}{c}
z_{\uparrow } \\ 
z_{\downarrow }%
\end{array}%
\right) =\frac{1}{\sqrt{|\eta _{\uparrow }|^{2p}+|\eta _{\downarrow }|^{2q}}}%
\left( 
\begin{array}{c}
\eta _{\uparrow }^{p} \\ 
\eta _{\downarrow }^{q}%
\end{array}%
\right) ,  \label{CP1}
\end{equation}%
where $\eta _{\uparrow }$ and $\eta _{\downarrow }$ are complex numbers
given by\cite{Duan,Moore, Deng,DengC,Ken} 
\begin{eqnarray}
\eta _{\uparrow }\left( \mathbf{k}\right) &=&\sin k_{x}+i\sin k_{y},  \notag
\\
\eta _{\downarrow }\left( \mathbf{k}\right) &=&\sin k_{z}+i(\cos k_{x}+\cos
k_{y}+\cos k_{z}-m),  \label{eta}
\end{eqnarray}%
while $p$ and $q$ are integers. We note that $p$ and $q$ are originally
introduced as a set of coprime integer\cite{Duan} but here we do not impose
it. We will see that $p$ and $q$ can be generalized even to rational numbers
though a cut is introduced. The normalized pseudospin is explicitly
represented as 
\begin{equation}
S_{x}+iS_{y}=\frac{2\eta _{\uparrow }^{p}\bar{\eta}_{\downarrow }^{q}}{|\eta
_{\uparrow }|^{2p}+|\eta _{\downarrow }|^{2q}},\quad S_{z}=\frac{\left\vert
\eta _{\uparrow }\right\vert ^{2p}-\left\vert \eta _{\downarrow }\right\vert
^{2q}}{|\eta _{\uparrow }|^{2p}+|\eta _{\downarrow }|^{2q}}.
\end{equation}%
We consider a class of pseudospin textures indexed by a pair of numbers $%
(p,q)$ in the Hamiltonian (\ref{HamilTH}).

The CP$^{1}$ field takes valued on the 3D sphere $\mathbb{S}^{3}$ since it
contains four real numbers, $N_{1}=\text{Re}\,z_{\uparrow }\left( \mathbf{k}%
\right) $, $N_{2}=\text{Im}\,z_{\uparrow }\left( \mathbf{k}\right) $, $N_{3}=%
\text{Re}\,z_{\downarrow }\left( \mathbf{k}\right) $ and $N_{4}=\text{Im}%
\,z_{\downarrow }\left( \mathbf{k}\right) $, together with the normalization
condition $\sum_{i}N_{i}^{2}=1$. It gives a mapping $\mathbb{T}%
^{3}\rightarrow \mathbb{S}^{3}$, for the Brillouin zone is a 3D torus. On
the other hand, the normalized pseudospin expressed as $\mathbf{S}=\mathbf{z}%
^{\dagger }\mathbf{\sigma z}$ defines a mapping $\mathbb{S}^{3}\rightarrow 
\mathbb{S}^{2}$, for it takes values on the sphere $\mathbb{S}^{2}$.
Consequently the underlying structure of the Hamiltonian (\ref{HamilTH}) is
a mapping $\mathbb{T}^{3}\rightarrow \mathbb{S}^{3}\rightarrow \mathbb{S}%
^{2} $ from the Brillouin zone to the pseudospin space.

The combined mapping $\mathbb{T}^{3}\rightarrow \mathbb{S}^{2}$ is indexed
by the Hopf number\cite{Moore,Deng,Duan,DengC,Ken,Xu}%
\begin{equation}
\Gamma (p,q)=-\frac{1}{2\pi ^{2}}\int_{\text{BZ}}d\mathbf{k\,}\varepsilon
_{\mu \nu \rho \tau }\,N_{\mu }\partial _{k_{x}}N_{\nu }\partial
_{k_{y}}N_{\rho }\partial _{k_{z}}N_{\tau }.  \label{HopfNum}
\end{equation}%
By inserting (\ref{eta}) to this formula we obtain%
\begin{equation}
\Gamma \left( p,q\right) =\left\{ 
\begin{array}{ccc}
0, &  & \left\vert m\right\vert >3 \\ 
pq, &  & 1<\left\vert m\right\vert <3 \\ 
-2pq, &  & \left\vert m\right\vert <1%
\end{array}%
\right. .  \label{GammaPQ}
\end{equation}%
We note that the definition of the Hopf number (\ref{HopfNum}) is different
from the previous literature\cite{Duan}, in which it is defined in terms of $%
(\eta _{\uparrow },\eta _{\downarrow })$ rather than $(z_{\uparrow
},z_{\downarrow })$. The formula (\ref{GammaPQ}) is understood intuitively
as follows: When $p=q=1$, it is a well-known formula for the Hopf number\cite%
{Duan}, which indicates that there exist one circle in the interior of the
torus and one circle around its axis of rotational symmetry. Now, $%
z_{\uparrow }\varpropto \eta _{\uparrow }^{p}$ and $z_{\downarrow
}\varpropto \eta _{\downarrow }^{p}$ imply that there exist $p$ and $q$ of
these circles.

The energy spectrum of the Hamiltonian (\ref{HamilTH}) reads $E\left( 
\mathbf{k}\right) =\pm \sqrt{S_{x}^{2}\left( \mathbf{k}\right)
+S_{y}^{2}\left( \mathbf{k}\right) +S_{z}^{2}\left( \mathbf{k}\right) }$.
The Fermi surface of the topological Hopf insulator is constructed by the
intersection of the three curved surfaces, $S_{x}=S_{y}=S_{z}=0$. In
general, the intersection of three surfaces is null, which results in an
insulating state.

In order to construct a model having a Fermi surface, it is necessary to
reduce the number of the conditions on the zero-energy states. There exist
three trivial ways to do so. We may employ any one of the conditions, $%
S_{x}=S_{z}=0$, $S_{y}=S_{z}=0$, or $S_{x}=S_{y}=0$. Obviously, they are the
zero-energy conditions of the following three Hamiltonians, 
\begin{eqnarray}
H_{xz}\left( \mathbf{k}\right) &=&S_{x}\left( \mathbf{k}\right) \sigma
_{x}+S_{z}\left( \mathbf{k}\right) \sigma _{z},  \label{HamilXZ} \\
H_{yz}\left( \mathbf{k}\right) &=&S_{y}\left( \mathbf{k}\right) \sigma
_{y}+S_{z}\left( \mathbf{k}\right) \sigma _{z},  \label{HamilYZ} \\
H_{xy}\left( \mathbf{k}\right) &=&S_{x}\left( \mathbf{k}\right) \sigma
_{x}+S_{y}\left( \mathbf{k}\right) \sigma _{y},  \label{HamilXY}
\end{eqnarray}%
respectively. The Fermi surface constructed by the intersection of the two
curved surfaces is a line node in general. In all these models we use the
same CP$^{1}$ field (\ref{CP1}) to define $S_{i}\left( \mathbf{k}\right) $.
Only the Hamiltonian $H_{xz}$ preserves the combination symmetry $PT$ of the
time reversal $T$ and inversion symmetry $P$.

\textit{The} $H_{xz}$ \textit{model}: We first investigate the Hamiltonian $%
H_{xz}$. From the normalization condition on $\mathbf{S}\left( \mathbf{k}%
\right) $, the zero-energy condition is expressed as $S_{y}\left( \mathbf{k}%
\right) =\pm 1$. Namely, the Fermi surface is the preimage of the points $%
\mathbf{S}_{\pm }=(0,\pm 1,0)$ in the combined Hopf map $\mathbb{T}%
^{3}\rightarrow \mathbb{S}^{2}$, which is a circle $\mathbb{S}^{1}$ in the
3D Brillouin zone. Consequently there are at least two loops corresponding
to the preimage of $S_{y}\left( \mathbf{k}\right) =1$ and $S_{y}\left( 
\mathbf{k}\right) =-1$ in the 3D Brillouin zone. These two loops form a link
when the Hopf number is nonzero.

\begin{figure}[t]
\centerline{\includegraphics[width=0.45\textwidth]{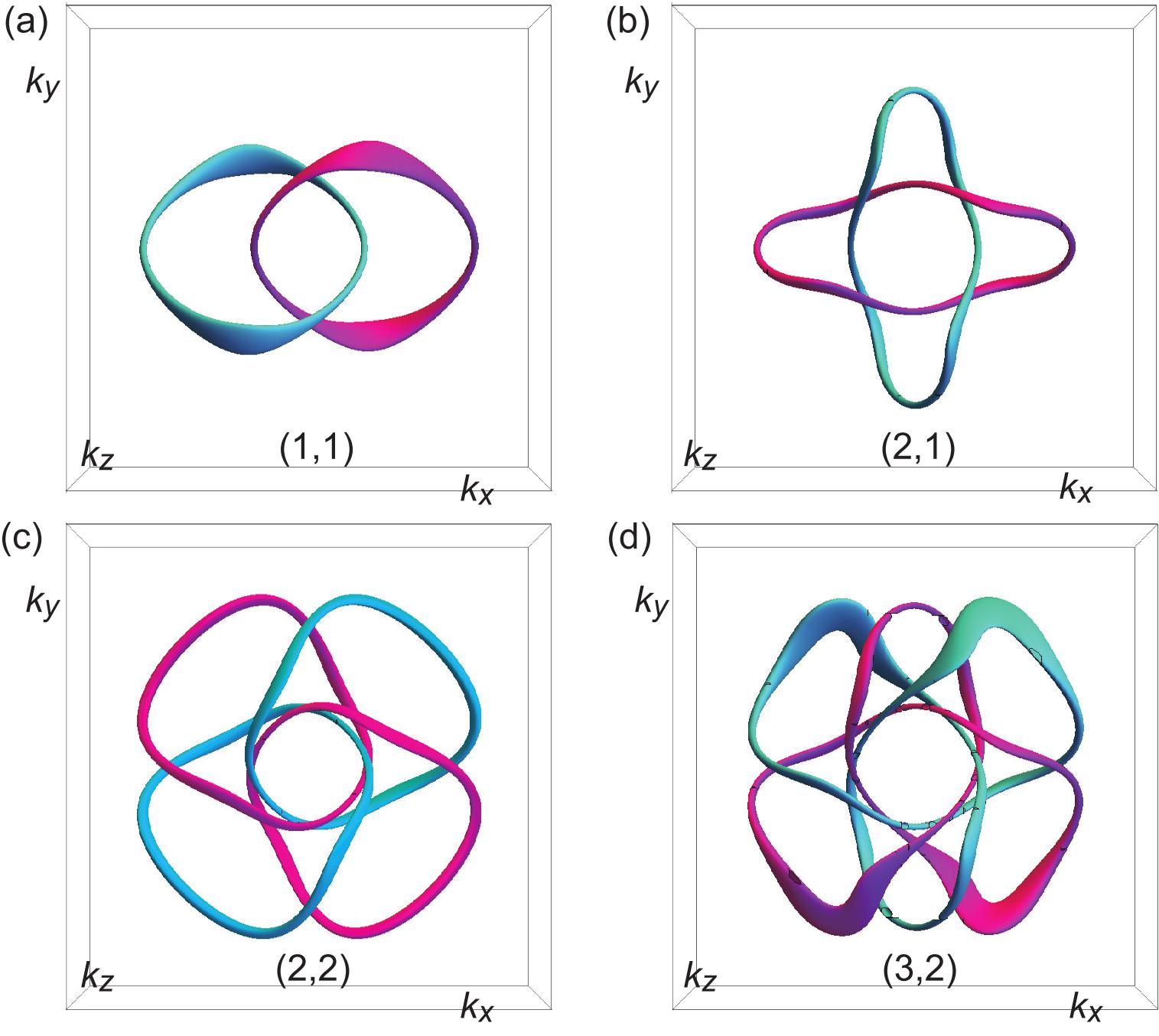}}
\caption{Top view of almost zero-energy surface of the Hamiltonian $H_{xz}$
in the 3D Brillouin zone. See the caption of Fig.\protect\ref{FigLink} for
(a)$\sim$(d).}
\label{FigKnot}
\end{figure}

Various links indexed by a pair $(p,q)$ are realized in the 3D Brillouin
zone. We show an almost zero-energy surface $E=\delta $ with $\left\vert
\delta \right\vert \ll 1$ for $(p,q)$ in Figs.\ref{FigLink} and \ref{FigKnot}%
. The preimage of $S_{y}\left( \mathbf{k}\right) =1$ is colored in magenta
and that of $S_{y}\left( \mathbf{k}\right) =-1$ is colored in cyan. For
example, the Hopf link and the Solomon's knot are realized by taking pairs $%
(1,1)$ and $(2,1)$, respectively. The Fermi surface for $(3,2)$ is given by
the combination of two trefoils, which we call a double trefoil.

We recall the terminology in the link theory. A link lying on the surface of
a torus is called a torus link. The torus link $T\left( p,q\right) $ winds $q
$ times around a circle in the interior of the torus, and $p$ times around
its axis of the rotational symmetry. In the present context, the surface
determined by the condition $S_{z}=0$ gives a torus. Thus the node indexed
by $\left( p,q\right) $ is the torus link $T\left( 2p,2q\right) $, where the
factor $2$\ appears because it is the preimage of the two points $S_{\pm
}=(0,\pm 1,0)$. According to the link theory, $T\left( 2p,2q\right) $ is
identical to $T\left( 2q,2p\right) $. Furthermore, $T\left( -2p,2q\right) $
link and $T\left( 2p,-2q\right) $ are mirror images of $T\left( 2p,2q\right) 
$.

If $p$\ and $q$\ are not relatively prime, we have a torus link with more
than one component. The number of the loops is given by gcd$\left(
2p,2q\right) =2$gcd$\left( p,q\right) $, where gcd represents the greatest
common divisor. For example, the Fermi surface consists of four loop nodes
for $p=2$ and $q=2$. This looks a bit odd since we have discussed that two
loops arise as the preimage of $S_{y}=\pm 1$. It is an interesting problem
how such a Fermi surface consisting more than two loops is realized for gcd$%
\left( p,q\right) \neq 1$. We assume gcd$\left( p,q\right) =s$. Then we can
write $p=sp^{\prime }$ and $q=sq^{\prime }$, where $p^{\prime }$ and $%
q^{\prime }$ are coprime integers. The solutions are given by $\eta
_{\uparrow }^{1/s}$ and $\eta _{\downarrow }^{1/s}$, where $\eta _{\uparrow }
$ and $\eta _{\downarrow }$ satisfy the $S_{y}=\pm 1$ for the model with $%
p^{\prime }$ and $q^{\prime }$. The roots of $\eta _{\uparrow }^{1/s}$ and $%
\eta _{\downarrow }^{1/s}$ have $s$ solutions, which results in the $s$
Fermi loops.

\begin{figure}[t]
\centerline{\includegraphics[width=0.45\textwidth]{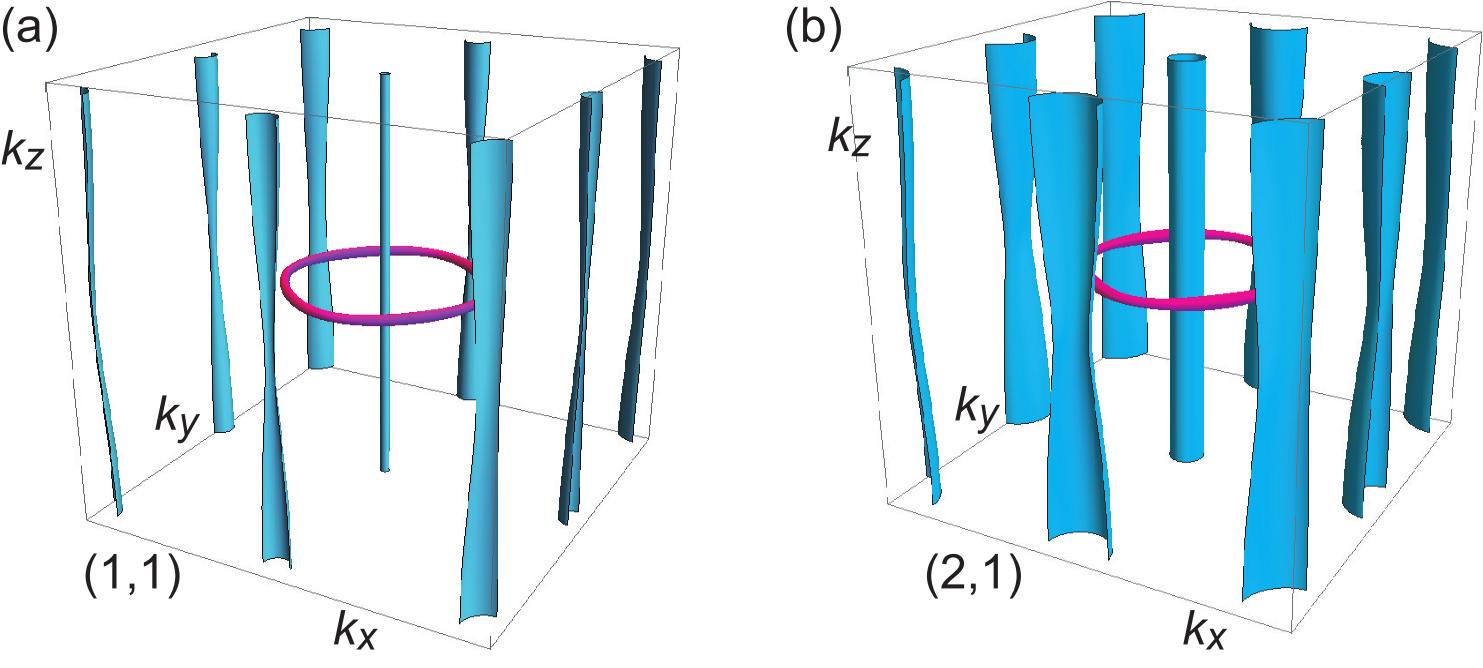}}
\caption{ Bird's eye's view of the almost zero-energy surface of the
Hamiltonian $H_{xy}$ for (a) the pair $(p,q)=(1,1)$ and (b) $(2,1)$. }
\label{FigKnotXY}
\end{figure}

In the following, we derive the equations to determine the link. The
zero-energy states must satisfy the condition $S_{z}=\mathbf{z}^{\dagger
}\sigma _{z}\mathbf{z}=\left\vert z_{\uparrow }\right\vert ^{2}-\left\vert
z_{\downarrow }\right\vert ^{2}=0$. By combining it with the normalization
condition $\left\vert z_{\uparrow }\right\vert ^{2}+\left\vert z_{\downarrow
}\right\vert ^{2}=1$, the CP$^{1}$ field is parametrized as 
\begin{equation}
z_{\uparrow }=\frac{1}{\sqrt{2}}e^{i\theta _{\uparrow }},\quad z_{\downarrow
}=\frac{1}{\sqrt{2}}e^{i\theta _{\downarrow }}.
\end{equation}%
The zero-energy states correspond to $S_{y}=\pm 1$, which reads 
\begin{equation*}
S_{y}=\mathbf{z}^{\dagger }\sigma _{y}\mathbf{z}=-\sin \left( \theta
_{\uparrow }-\theta _{\downarrow }\right) =\pm 1.
\end{equation*}%
The solution is given by $\theta _{\uparrow }-\theta _{\downarrow }=\mp 
\frac{\pi }{2}$. We find the relation $z_{\downarrow }=\pm iz_{\uparrow }$,
or%
\begin{equation}
\eta _{\downarrow }^{q}=\pm i\eta _{\uparrow }^{p}.
\end{equation}%
These two equations determine the Fermi surface made of links.

\textit{The} $H_{yz}$ \textit{model}: The Fermi surface of $H_{yz}$ is
topologically equivalent to that of $H_{xz}$ model. The only difference is
that the Fermi surface is rotated 90 degree between the $H_{xz}$ and $H_{yz}$
models.

\textit{The} $H_{xy}$ \textit{model}: The Fermi surface of $H_{xy}$ looks
very different, where some of the Fermi surfaces form lines penetrating the
whole Brillouin zone: See Fig.\ref{FigKnotXY}. The model $H_{xy}$ is
instructive since we obtain various analytical expressions.

The zero-energy condition in the model $H_{xy}$ is given by the preimage of $%
S_{z}=\left\vert z_{\uparrow }\right\vert ^{2}-\left\vert z_{\downarrow
}\right\vert ^{2}=\pm 1$. By combining it with the normalization condition $%
\left\vert z_{\uparrow }\right\vert ^{2}+\left\vert z_{\downarrow
}\right\vert ^{2}=1$, the preimage of $S_{z}=1$ is given by $\left\vert
z_{\uparrow }\right\vert ^{2}=1$ and $\left\vert z_{\downarrow }\right\vert
^{2}=0$, which is equivalent to the condition $\left\vert z_{\downarrow
}\right\vert ^{2}=\left\vert \eta _{\downarrow }\right\vert ^{2q}=0$, where%
\begin{equation}
\left\vert \eta _{\downarrow }\right\vert ^{2}=\sin ^{2}k_{z}+\left( \cos
k_{x}+\cos k_{y}+\cos k_{z}-m\right) ^{2}=0.
\end{equation}%
The solution is 
\begin{equation}
\cos k_{x}+\cos k_{y}=m-1,\quad k_{z}=0,  \label{LoopXY}
\end{equation}%
which represents $q$-fold degenerate loop nodes.

\begin{figure}[t]
\centerline{\includegraphics[width=0.45\textwidth]{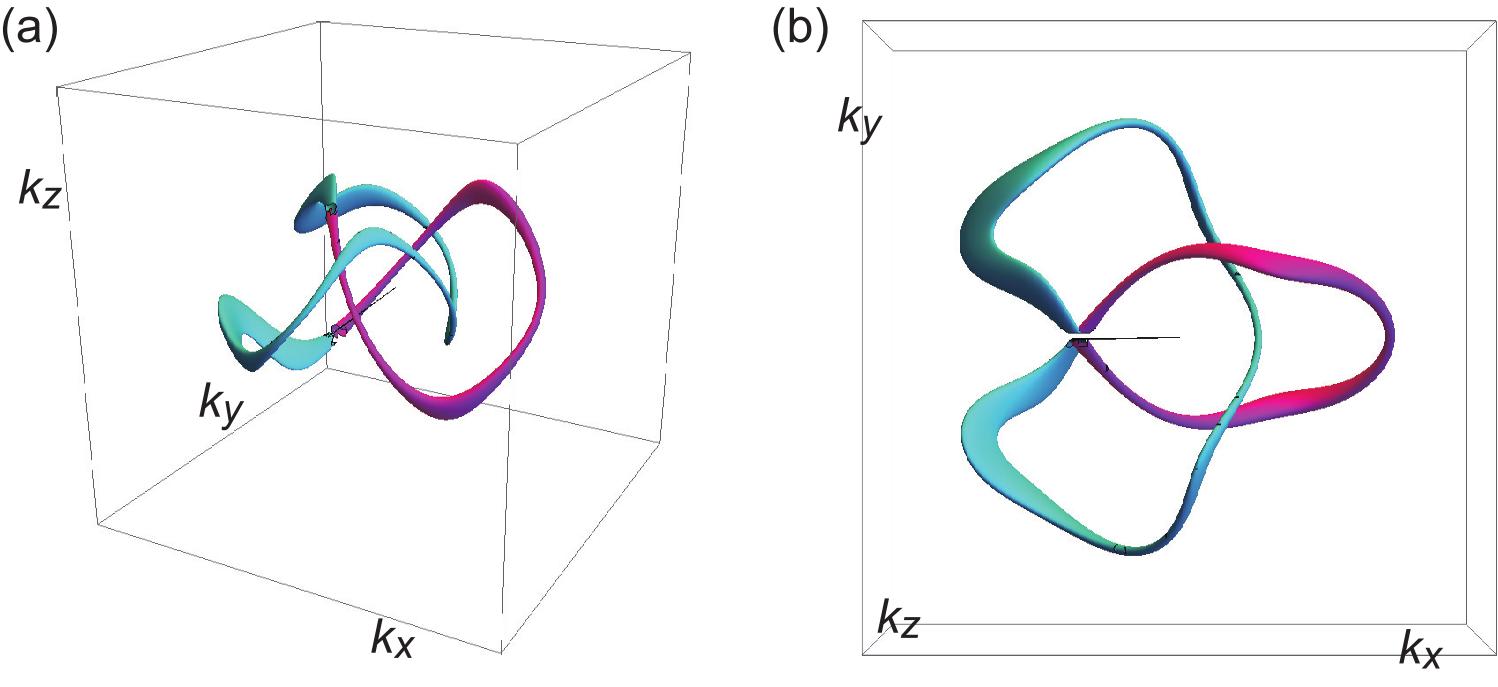}}
\caption{Trefoil-knot semimetal with $p=3/2$ and $q=1$. (a) Bird's eye's
view and (b) the top view of the almost zero-energy surface. There is a cut
on the $k_y=0$ plane for $k_x<0$, where the magenta and cyan curves touch.}
\label{FigTrefoil}
\end{figure}

On the other hand, the preimage of $S_{z}=-1$ is given by $\left\vert
z_{\uparrow }\right\vert ^{2}=0$ and $\left\vert z_{\downarrow }\right\vert
^{2}=1$, which is equivalent to the condition $\left\vert z_{\uparrow
}\right\vert ^{2}=\left\vert \eta _{\uparrow }\right\vert ^{2p}=0$, where%
\begin{equation*}
\left\vert \eta _{\uparrow }\right\vert ^{2}=\sin ^{2}k_{x}+\sin ^{2}k_{y}=0.
\end{equation*}%
The solution represents p-fold degenerate line nodes along the $k_{z}$
direction described by $k_{x}=n_x\pi $, $k_{y}=n_y\pi $, $k_{z}$ with $-\pi
\leq k_{z}\leq \pi $, where $n_x$ and $n_y$ are integers.

We may also analyze this Hamiltonian system from the view point of the Berry
curvature. We introduce a continuum model defined by%
\begin{eqnarray}
\eta _{\uparrow }\left( \mathbf{k}\right) &=&k_{x}+ik_{y},  \notag \\
\eta _{\downarrow }\left( \mathbf{k}\right) &=&k_{z}+i(3-m-\left(
k_{x}^{2}+k_{y}^{2}+k_{z}^{2}\right) /2).
\end{eqnarray}%
The Fermi surface is topologically equivalent to the original model (\ref%
{eta}). The only difference is the approximation of the loop (\ref{LoopXY})
by the circle $k_{x}^{2}+k_{y}^{2}=2(3-m)$, which does not change the
linking structure.

\begin{figure}[t]
\centerline{\includegraphics[width=0.45\textwidth]{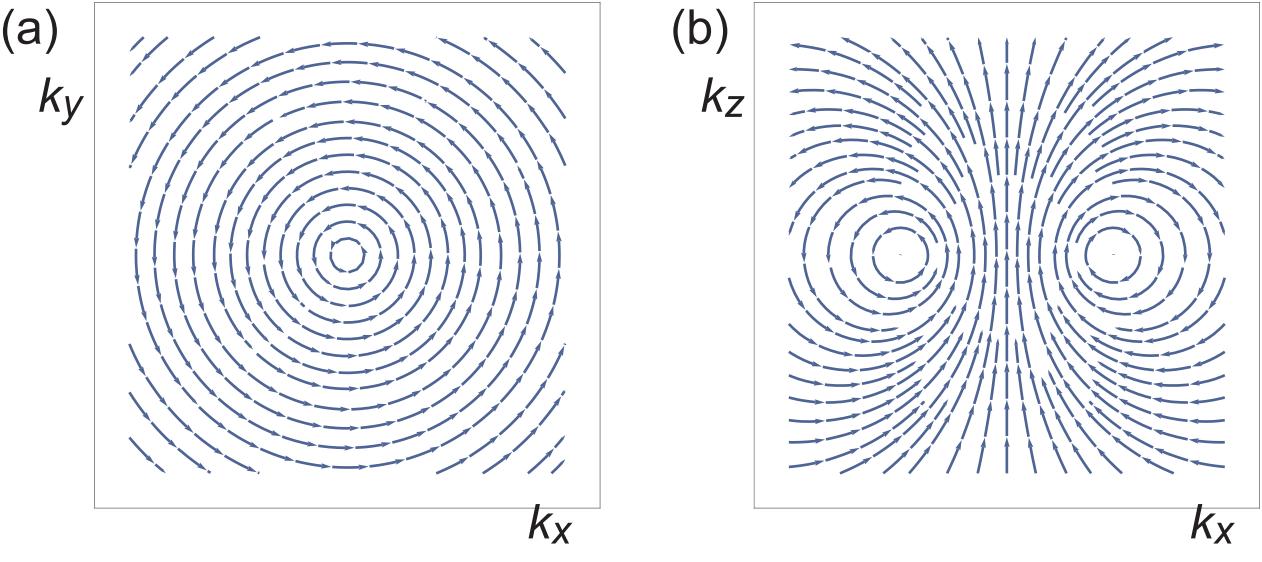}}
\caption{ (a) Stream plot of the Berry curvature ($A_{x}$, $A_{y}$) along
the $k_{x}$-$k_{y}$ plane with $k_{z}=0$ and (b) Stream plot of the Berry
curvature ($A_{x}$, $A_{z}$) on the $k_{x}$-$k_{z}$ plane at $k_{y}=0$.}
\label{FigBerry}
\end{figure}

By using them we may derive explicitly the eigenstate $\left\vert \psi
\right\rangle $ of the Hamiltonian (\ref{HamilXY}). The Berry connection $%
A_{k_{i}}=-i\left\langle \psi \right\vert \partial _{k_{i}}\left\vert \psi
\right\rangle $ is given by%
\begin{eqnarray}
A_{k} &=&\frac{2kqk_{z}}{k_{z}^{4}+k_{z}^{2}\left( k^{2}+2m-4\right) +\left(
k^{2}+2m-6\right) ^{2}},\quad \\
A_{\theta } &=&\frac{p}{2}, \\
A_{z} &=&\frac{q\left( k_{z}^{2}-k^{2}-2m+6\right) }{k_{z}^{4}+k_{z}^{2}%
\left( k^{2}+2m-4\right) +\left( k^{2}+2m-6\right) ^{2}},
\end{eqnarray}%
where we have introduced the polar coordinate $k_{x}=k\cos \theta $ and $%
k_{y}=k\sin \theta $. We show the Berry curvature in Fig.\ref{FigBerry}. We
find a vortex structure along a line node (cyan line in Fig.\ref{FigKnotXY})
described by $k_{x}=0$, $k_{y}=0$, $k_{z}$ with $-\pi \leq k_{z}\leq \pi $,
and a circle described by $k_{x}^{2}+k_{y}^{2}=6-2m$, $k_{z}=0$. Actually
this line node has a p-fold degeneracy. Indeed, we calculate the Berry phase
along a loop encircling the line node to find that 
\begin{equation}
\Gamma _{\text{B}}=\oint A_{i}dk_{i}=\int_{0}^{2\pi }A_{\theta }d\theta
=p\pi ,
\end{equation}%
which is quantized and represents the degeneracy. In the same way, the Berry
phase along a small circle around a loop node (magenta loop in Fig.\ref%
{FigKnotXY}) is obtained as 
\begin{equation}
\Gamma _{\text{B}}=\oint A_{i}dk_{i}=\int_{0}^{2\pi }A_{\theta }d\theta
=q\pi ,
\end{equation}%
representing the degeneracy of the loop node. Thus the indices $p$ and $q$
are detected separately, while only the product $pq$ appears in the Hopf
number.

\begin{figure}[t]
\centerline{\includegraphics[width=0.45\textwidth]{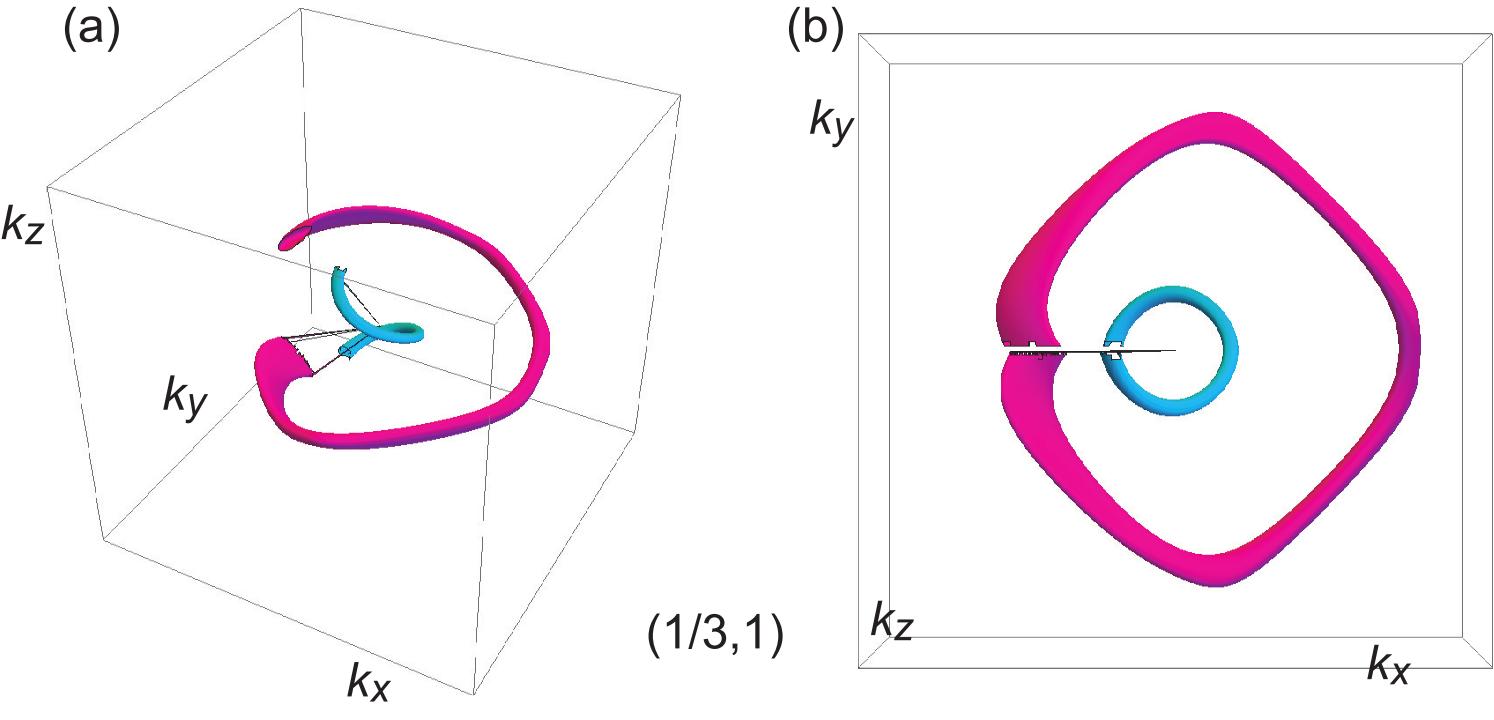}}
\caption{ Open-string semimetal with $p=1/3$ and $q=1$. (a) Bird's eye's
view and (b) top view of almost zero-energy surface. There is a cut along
the $k_{y}=0$ plane for $k_{x}<0$, where the ends of the open string exist.}
\label{FigString}
\end{figure}

\textit{Torus-knot semimetals:} We have seen that the Fermi surfaces consist
of at least two loops, which form a torus link. It is an interesting problem
whether we can construct a Fermi surface consist of one nontrivial loop,
which is a torus knot. Since the number of loops is given by $2$gcd$(p,q)$,
we must take one of $p$ and $q$ non-integer. We find that a torus-knot Fermi
surface is realized by taking a half-integer $p$. For example, we can
realize a trefoil Fermi surface by taking $p=3/2$ and $q=1$, which is shown
in Fig.\ref{FigTrefoil}. There is a cut in the momentum space due to the
square root contribution in the Hamiltonian. Namely, only the magenta or
cyan curve does not form a closed loop but form a loop with the combination
of them.

\textit{Open string semimetals:} It is possible to choose even any rational
numbers for $p$ or $q$, which creates an open string Fermi surface as shown
in Fig.\ref{FigString}. This is understood as follows. The model with $p$
and $q$ gives the torus link $T\left( 2p,2q\right) $. If one of the $p$ and $%
q$ is not a half integer, it cannot describe a closed loop, generating to an
open string. The ends of the open string are on the cut plane.

The author is very much grateful to N. Nagaosa for many helpful discussions
on the subject. This work is supported by the Grants-in-Aid for Scientific
Research from MEXT KAKENHI (Grant Nos.JP17K05490 and 15H05854).

\textit{Note added:} During the preparation of this manuscript, we became
aware of closely related works\cite{WChen,ZYan,PYChang}, where various
linked nodal semimetals are proposed. Especially, this work has turned out
to be a generalization of the work \cite{ZYan}, where only the case with $%
p=q=1$ is studied.


\end{document}